\pdfoutput=1

\documentclass[11pt]{article}

\usepackage[final]{acl}

\usepackage{times}
\usepackage{latexsym}

\usepackage[T1]{fontenc}

\usepackage[utf8]{inputenc}

\usepackage{microtype}

\usepackage{inconsolata}

\usepackage{enumitem}
\usepackage{booktabs}
\usepackage{caption}
\usepackage{subcaption}
\usepackage{booktabs}
\usepackage{multirow}
\usepackage{amssymb}
\usepackage{amsmath}
\usepackage{float}
\usepackage{graphicx}
\DeclareUnicodeCharacter{1D54F}{$\mathbb{X}$}

\title{$\mathbb{X}$-posing Free Speech: Examining the Impact of Moderation Relaxation on Online Social Networks}

\author{Arvindh Arun\thanks{Equal Contribution.} \\
  IIIT, Hyderabad\\
  \texttt{arvindh.a@research.iiit.ac.in} \\\And
  Saurav Chhatani\footnotemark[1] \\
  IIIT, Hyderabad\\
  \texttt{saurav.chhatani@students.iiit.ac.in}\\\AND
  Jisun An \\
  Indiana University, Bloomington\\
  \texttt{jisunan@iu.edu} \\ \\\And
  Ponnurangam Kumaraguru \\
  IIIT, Hyderabad \\
  \texttt{pk.guru@iiit.ac.in}}

\begin{document}
\maketitle
\begin{center}
    \textbf{\textcolor{red}{WARNING}\\ The following text contains offensive words}
\end{center}
\begin{abstract}
We investigate the impact of free speech and the relaxation of moderation on online social media platforms using Elon Musk's takeover of Twitter as a case study. By curating a dataset of over 10 million tweets, our study employs a novel framework combining content and network analysis. Our findings reveal a significant increase in the distribution of certain forms of hate content, particularly targeting the LGBTQ+ community and liberals. Network analysis reveals the formation of cohesive hate communities facilitated by influential bridge users, with substantial growth in interactions hinting at increased hate production and diffusion. By tracking the temporal evolution of PageRank, we identify key influencers, primarily self-identified far-right supporters disseminating hate against liberals and woke culture. Ironically, embracing free speech principles appears to have enabled hate speech against the very concept of freedom of expression and free speech itself. Our findings underscore the delicate balance platforms must strike between open expression and robust moderation to curb the proliferation of hate online.
\end{abstract}

\section{Introduction}
Social media platforms have become the primary forum for public discussion in today's digital world. While this surge of content fosters a diversity of viewpoints, it also presents significant challenges in maintaining a healthy, productive, and inclusive online environment. One of the most pressing issues is the detection and management of abusive content~\citep{lenhart2016online,davidson2017automated}. Traditional moderation methods, reliant on automated systems and centralized teams, are increasingly struggling to keep pace with the ever-growing content volume and the complex nature of abusive content~\citep{pavlopoulos2020toxicity}.
In response, community moderation, also known as crowd moderation, has emerged as a promising strategy for safeguarding online spaces~\citep{cullen2022practicing,lampe2014crowdsourcing,seering2023moderates}. This approach leverages the collective vigilance of community members, empowering them to participate directly in content moderation~\citep{matias2019preventing}.

Current research primarily focuses on the effects of stricter moderation practices, such as account bans or subreddit closures~\citep{ali2021understanding,chandrasekharan2022quarantined,horta2021platform}, and their impact on user behavior and community dynamics~\citep{cheng2015antisocial}. While these studies offer valuable insights into online control mechanisms, a gap exists in our understanding of how loosening community moderation impacts user behavior and discourse dynamics. Examining how reduced moderation intensity shapes online community norms and interactions is crucial, as it can offer nuanced perspectives on striking the right balance between enabling free expression and upholding community standards within digital spaces.

Our work aims to address this gap by exploring the ramifications of diminished moderation within online communities. 
Recently, the push for free speech and the voices supporting the downfall of heavy moderation have been resonant \citep{woah2022}. 
Many platforms, including $\mathbb{X}$ (henceforth referred to as Twitter), have opted to relax their moderation policies and open-sourced their algorithms for transparency.
Elon Musk's infusion of Free Speech on Twitter could unleash a flurry of support for similar measures in other platforms due to the shifting societal norms and the onset of the woke culture that emphasizes inclusivity and diverse perspectives~\citep{woke}, potentially leading platforms to adapt their rules and practices to align with these evolving norms. 
Previous studies have documented a rise in hate speech and bot activity subsequent to the acquisition of Twitter~\citep{auditing,spikes}. However, research has yet to explore how this takeover (moderation relaxation) has impacted community engagement dynamics.

By integrating content and network analysis, our work probes into the shifts in linguistic patterns and user interactions on Twitter, thoroughly exploring how free speech and hate content propagation intertwine following Elon Musk's takeover.
Specifically, we focus on three primary research questions,
\begin{enumerate}[noitemsep,topsep=0pt]
    \item How did the Hate Speech landscape change after the relaxation of moderation?
    \item How does the moderation relaxation affect the hate in existing communities?
    \item Can we (early) detect the users who drove the change in this landscape?
\end{enumerate}

\section{Related Work}
\textbf{Hate Speech and Moderation.}
\citet{gab} report that Gab, a platform designed as a less restrictive alternative to Twitter, had a higher prevalence of hate speech attributed to its appeal among alt-right users, conspiracy theorists, and those with extremist views. 
These findings highlight the potential risks associated with loosening moderation policies.
Prior studies demonstrate how platform-wide moderation interventions, such as account bans or subreddit quarantines, can effectively mitigate hate speech and disrupt the growth of harmful communities~\citep{reddit-ban, chandrasekharan2022quarantined}.

\noindent\textbf{Hateful User Detection.} To detect hateful users, \citet{user-representation} propose a model that uses intra-user representation learning on a user's historical posts and inter-user representation learning across similar posts by other users. \citet{early-detection} find that hateful user detection performance increases by combining BOW models with user-level representations based on latent author topics and user embeddings. \citet{characterizing} emphasize the challenges of hate speech detection due to the subjectivity and noise inherent in social media text. Thus, activity patterns, word usage, and network structure are used to detect hateful users. Also, \citet{brutus} demonstrates significant performance improvements when both textual features and social connections are used.

\section{Dataset}
Existing research on the effects of Elon Musk's takeover of Twitter mainly measures surface-level metrics like volume of hate speech and bot activity or are comparative studies on older datasets~\citep{dataset, auditing}. Yet, understanding the impact on user interactions, community formation, and influential user interactions necessitates data on network dynamics, which current datasets do not provide. To address this, we curate a new dataset\footnote{The dataset can be shared upon a formal request} that tracks hate speech and models user interactions surrounding this content.

\subsection{Hateful tweet extraction (D1)}
\label{sec:d1}

Following~\citet{ethnic-slurs}, we first collate a list of ethnic slurs from Wikipedia.\footnote{\href{https://en.wikipedia.org/wiki/List_of_ethnic_slurs}{Wikipedia List of Ethnic Slurs}} From this list, we manually selected a subset (henceforth referred to as keywords) based on various factors such as their severity, relevance on social media platforms, and diversity. To further refine the keywords suitable for our analysis, we conducted a trial run by querying these keywords on Twitter's Academic API for a few days, noting their relative frequency and relevance to the scope of our study. After this filtering process, we converged on the final set of 32 keywords to be used for data collection.

We set the timeline in focus containing a month before the takeover (Sept. 27 to Oct. 27, 2022), the day of the public announcement of the takeover (Oct. 28, 2022), and a month after it (Oct. 29 to Nov. 28, 2022).

We use the Academic API for data collection, aiming to collect relevant data exhaustively and not just a representative subsample. Our script loops day-by-day for all 63 days and collects all tweets satisfying the following conditions:
(1) Language labeled as EN, (2) Not a Retweet, and (3) Contains at least one of the keywords. The collection is exhaustive as we impose no hard limit on the quantity. We collect 1,008,111 tweets posted by 584,416 unique users whose cumulative retweet count is 886,162.

\subsection{Hateful user timeline collection (D2)}
\label{sec:d2}
We also collect user-specific tweets to identify users who drive significant change. For this, we explore several hate classification models to apply another stricter filter over the collected dataset to improve its quality. Following~\citet{hatexplain}, we use HateXplain\footnote{\href{https://huggingface.co/Hate-speech-CNERG/bert-base-uncased-hatexplain-rationale-two}{Hate-speech-CNERG/bert-base-uncased-hatexplain-rationale-two}} which outputs a probability value between 0 and 1 for each tweet, with 0 being Normal and 1 being Abusive. As it is a probability distribution, we take 0.5 as the default threshold. To justify the threshold, we manually annotate 200 tweets and compare them with the score given by HateXplain, verifying that a threshold is ideal by inter-annotator agreement. We subsequently filter 288,566 gold-standard hateful tweets, of which 87,027 have at least 1 Retweet in the dataset.

Focusing on 6,168 users who posted at least three hateful tweets (i.e., key contributors) out of a total of 202,884, we collect their entire Twitter activity in our chosen time of focus, including original tweets, replies, quotes, and retweets. This leads to a total collection of 9,716,185 tweets, of which 1,772,072 are original tweets. 

\section{Experiments}
\subsection{How did the Hate Speech landscape change?}
To study the changes in the linguistic landscape, we analyze 1) Types of hate speech, 2) Representative words, and 3) Shifts in word semantics.  

\begin{table}[]
\centering
\caption{Percentage increase in the composition of Hateful content by category}
\label{tab:category-table}
\begin{tabular}{|c|c|}
\hline
\textbf{Hate category}             & \textbf{\% Increase (p-value)} \\
\hline 
Sexism                         & 22.8 (<0.0001)\\
\hline 
Racism                         & 50.5 (<0.0001)\\
\hline 
Disability                     & 53.3 (0.0019)\\
\hline 
Sexual Orientation             & 38.6 (<0.0001) \\
\hline 
Religion                       & 50.2 (<0.0001)\\
\hline 
Other                          & 16.4 (0.0007)\\
\hline
\end{tabular}
\end{table}

\subsubsection{Types of Hate Speech}
\label{sec:categorization}
We first analyze the changes in the prevalence of the keywords in D1 after the moderation relaxation. We find a 32.81\% increase in tweets containing the selected keywords. The keywords with the most significant increases are the `n****r' by 83.3\%, `d*rkie' by 81.1\%, `com*ie' by 64.7\%, `h*lf-br**d' by 43.6\%, and `paj**t' by 35.7\%. The prevalent use of such terms in the dataset post-relaxation implies a significant rise in certain forms of hate speech, reflecting the shifting landscape on Twitter.

To gain deeper insights, we analyze the specific categories of hate speech that became more prevalent after the relaxation. For categorizing hate tweets, we evaluate several popular open-source models and select \textit{Twitter Roberta Base Hate Multiclass} \citep{roberta},\footnote{\href{https://huggingface.co/cardiffnlp/twitter-roberta-base-hate-multiclass-latest}{cardiffnlp/twitter-roberta-base-hate-multiclass-latest}} as it is trained on a diverse corpus of tweets compiled from thirteen distinct datasets, making it highly relevant and well-suited for our study. Moreover, this model performs best when manually verified on a subset of tweets from our dataset. It is trained to classify each tweet into one of several categories: sexism, racism, disability hate, hate based on sexual orientation, religious hate, other types of hate, or non-hate speech.

We preprocess the tweets by converting the text to lowercase, removing mentions, non-alphabetic characters, and URLs before feeding them through the categorization model. Our analysis focuses specifically on the original tweets posted by the identified hateful users (D2), excluding replies, retweets, and quoted tweets, providing a clearer insight into their patterns of hate speech without the noise of external interactions.

Overall, we note a significant 32.6\% rise in the hate speech composition on D2 post-relaxation. Table \ref{tab:category-table} shows the category-wise percentage increase where all categories see an increase in their composition, with the most being in Disability (53.3\%), Religion (50.2\%), and Racism (50.5\%) with low p-values confirming their statistical significance.

\subsubsection{Representative Words}
\label{sec:log-odds}

To identify shifts in language tone and term usage across categories, we employed log-odds ratios combined with informative Dirichlet priors and word frequency analysis, following \citet{zscore}. We employ the same preprocessing steps detailed in \ref{sec:categorization}. Additionally, we lemmatize words for uniformity and exclude words under three characters to improve data quality. We then calculated the log-odds ratio (z-score) for each word between the pre and post-takeover corpora, using prior frequencies from the Google Books Ngram corpus \citep{ngram}.\footnote{\href{https://storage.googleapis.com/books/ngrams/books/datasetsv2.html}{Google Books Ngram Viewer}}  This method identifies representative words unique to each corpus based on significance within each. Words were filtered based on both z-score and frequency, selecting the top 50 for the pre-takeover corpus and the bottom 50 for the post-takeover corpus, with a minimum frequency threshold of 1\% of their respective corpus size.

Examining the category of \textit{disability}, before the relaxation, prevalent words such as `retarded', `f**k', and `stupid' underscore a pervasive use of derogatory language. Post-relaxation, these terms persist, joined by others like `tra*ny' and `schizo', further stigmatizing individuals with mental health conditions and transgender identity.

\begin{table}[]
\centering
\caption{Cosine similarity association of Topics and Keywords before and after the relaxation}
\label{tab:context-table}
\begin{tabular}{@{}ccc@{}}
\toprule
\textbf{(Topic, Keyword)}                      & \textbf{Before} & \textbf{After} \\ \midrule
\multirow{1}{*}{(Moderation, Curbing)}                      & 0.02              & 0.37             \\ \midrule
\multirow{1}{*}{(Free Speech,}           & <0.001                 & 0.34             \\
\multirow{1}{*}{Elonmuskbuystwitter)}           &                 &              \\
\multirow{1}{*}{(Free Speech, Facebooknazis)}           & <0.001                 & 0.32             \\ \midrule
\multirow{1}{*}{(Hate Speech, Free)}                         & 0.22              & 0.39             \\
\multirow{1}{*}{(Hate Speech, Facebooknazis)}                & <0.001              & 0.34             \\
\multirow{1}{*}{(Hate Speech, Unrestricted)}                & 0.15              & 0.33             \\
\multirow{1}{*}{(Hate Speech, Neonazis)}                & 0.14              & 0.31             \\\midrule
\multirow{1}{*}{(Liberal, Commie)}                & 0.26              & 0.32             \\
\multirow{1}{*}{(Liberal, Worktard)}                & <0.001              & 0.38             \\
\multirow{1}{*}{(Liberal, Millionvotesmyass)}                & <0.001              & 0.38             \\
\multirow{1}{*}{(Liberal, Fakeelection)}                & <0.001              & 0.32             \\
\multirow{1}{*}{(Liberal, Bidensucks)}                & <0.001              & 0.32             \\
\multirow{1}{*}{(Liberal, MAGA)}                & <0.001              & 0.30             \\
\multirow{1}{*}{(Liberal, Womansplaining)}                & 0.17              & 0.31      \\
\midrule
\multirow{1}{*}{(Conservative, Fuckthegop)}                & <0.001              & 0.32             \\
\multirow{1}{*}{(Conservative, Semifacist)}                & <0.001              & 0.29             \\ \midrule
\multirow{1}{*}{(Woke, Wokeisdead)}                & <0.001              & 0.28 
\\ 
\multirow{1}{*}{(Woke, Babykilling)}                & <0.001              & 0.34
\\ \bottomrule                             
\end{tabular}
\end{table}

In the category of \textit{racism}, pre-relaxation terms like `com*ie' and `slave' targeted specific ethnic or political groups. Post-relaxation, there was a marked increase in the usage of highly offensive racial slurs like `n***a' and `n****r'. Before the relaxation, terms such as `chinese' and `illegal' hinted at racial discrimination against specific ethnic or immigrant groups. Post-relaxation, a focus on racial and political divisions emerged through terms like `black' and `democrat', accompanied by a surge in explicit language, reflecting a shift towards more vulgar expressions of racism.

In the category of \textit{religion}, post-relaxation discourse intensified with terms like `murderous' and `evil', signaling a move towards more extreme and negative portrayals of religious concepts.

However, for categories \textit{sexism}, \textit{sexual orientation}, and \textit{other}, our analysis didn't reveal a significant shift in representative words following the takeover.

\begin{figure*}[]
     \centering
     \begin{subfigure}{.45\textwidth}
         \centering
         \includegraphics[width=\linewidth]{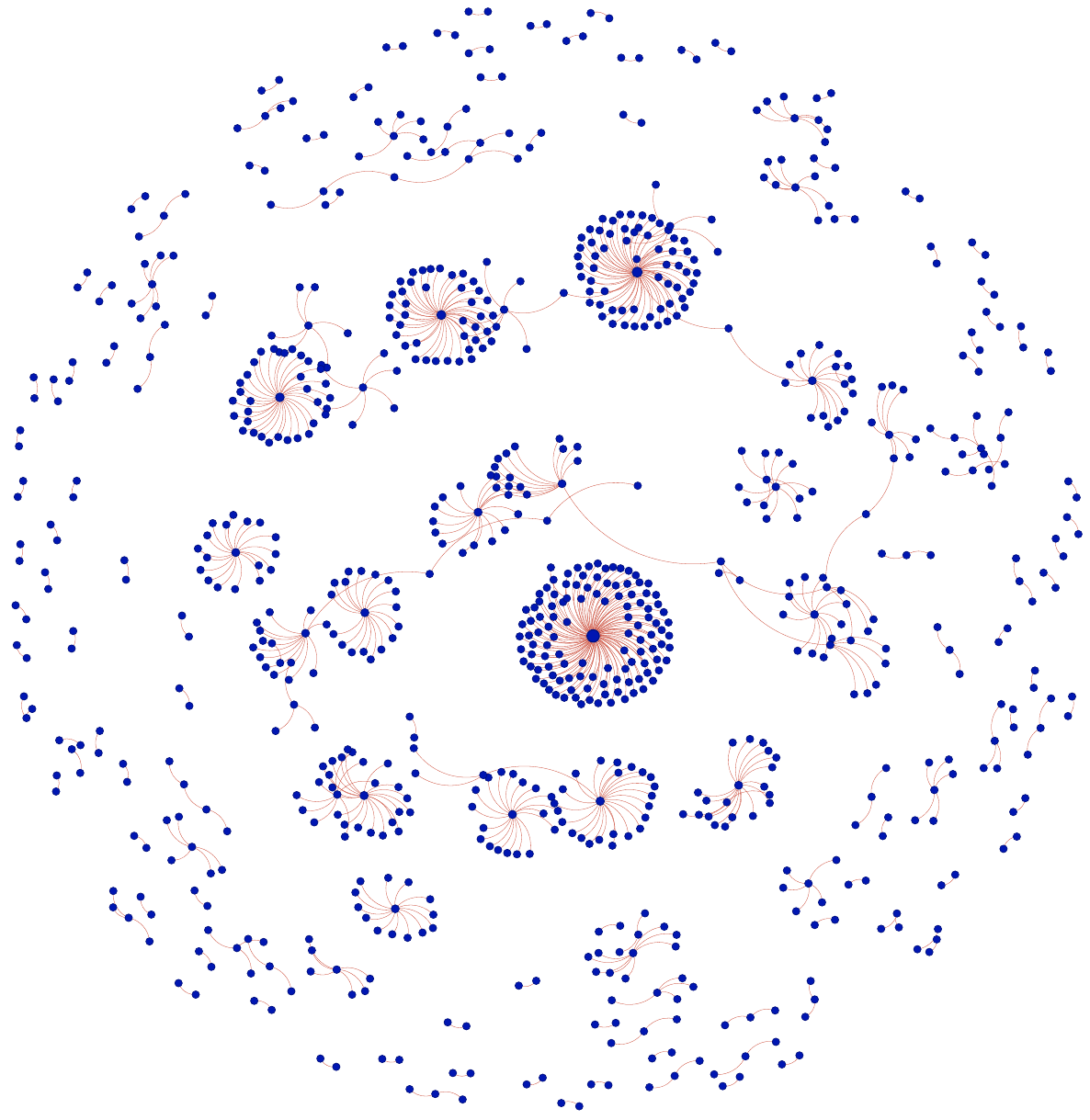}
         \caption{2 weeks before the takeover}
         \label{fig:start}
     \end{subfigure}
     \hspace{5mm}
     \begin{subfigure}{.45\textwidth}
         \centering
         \includegraphics[width=\linewidth]{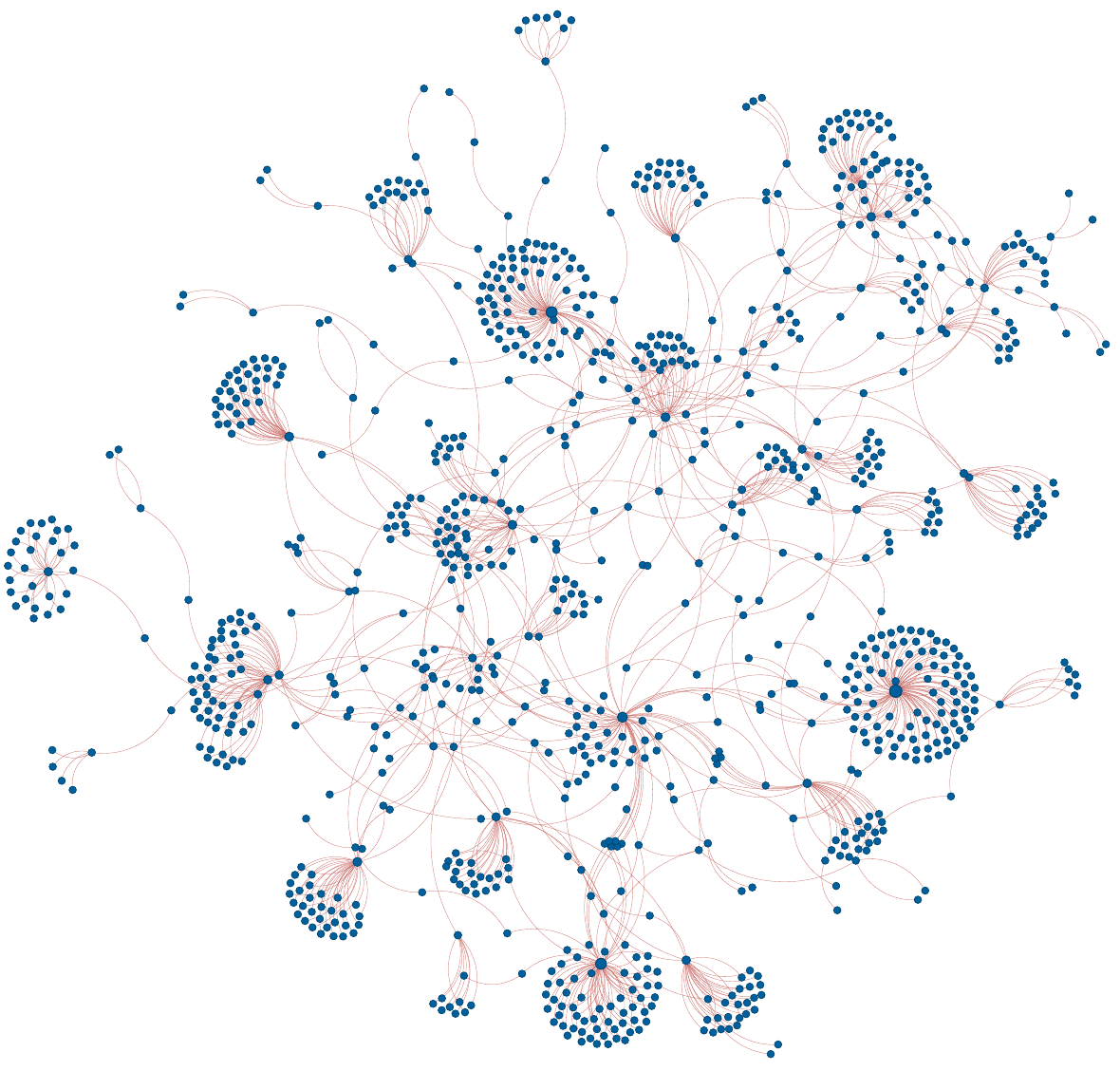}
         \caption{2 weeks after the takeover}
         \label{fig:final}
     \end{subfigure}
    \caption{ForestFire subsampled $(|V|=1000)$ visualization of hate interaction network two weeks before and two weeks after the takeover}
    \vspace{-5mm} 
    \label{fig:types}
\end{figure*}

\subsubsection{Shifts in Word Semantics}
\label{sec:w2v}
Finally, we analyze shifts in semantics to identify entities increasingly associated with hate speech after the relaxation. To investigate changes in word semantics, we employ a word2vec model, trained separately on datasets from each timeline on D2 (all user tweets published before and after the takeover), following \citet{w2v}.
This approach is based on the premise that words frequently used together in sentences will be positioned closer to the model's latent space. By examining these spatial relationships, we aim to identify significant contextual shifts of words after the relaxation.

Analyzing contextual changes associated with keywords reveals increased hate towards political agendas, particularly the left wing. Notably, the irony arises as the left's advocacy for free speech intensifies, yet our results indicate an increased critique against these left-wing agendas. Table \ref{tab:context-table} illustrates the cosine similarity between the topic and the keyword before and after Elon Musk's takeover. The increased association between \textit{Moderation} and \textit{Curbing} suggests discussions on decreased moderation. The term \textit{Facebooknazis}, critiquing strict moderation on Facebook, becomes closely linked with  \textit{Hate Speech} and \textit{Free Speech}. \textit{Elonmuskbuystwitter} shows a strong association with free speech, reflecting the impact of Elon Musk's takeover in this context. The rise in association between \textit{Hate Speech} and \textit{Free} suggests perceived liberalization enabling more hateful content circulation. \textit{Liberal} and \textit{Conservative} are associated more with negative terms post-relaxation, indicating heightened political polarization. Increased association of \textit{Liberal} with extreme right-wing terms like \textit{MAGA} signifies a stronger pro-Trump presence post-relaxation. \textit{Woke} also becomes more associated with \textit{Wokeisdead} suggesting increased hostility.

\subsection{How does the moderation relaxation affect the hate in existing communities?}
\label{sec:rq2}
To understand the evolution of hate communities and user behavior, we construct the most representative interaction network between the users. 
As previous studies have shown that retweets on Twitter are the most representative of homophilic interactions \citep{homophily}, using D1, we construct a retweet interaction network of the 7,385 hateful tweets having at least one retweet that the above-chosen 6,168 users have posted.

Due to the absence of explicit timestamps in the Twitter API for retweets, we discretize time intervals into 40 days, aligning with the observation that a significant portion of retweets occurs within the same day as the original tweet \citep{retweet-time}, resulting in a network that grows each day for the entire period. We also explore various versions of the construction, like considering each timestamp's incoming edges as separate networks, adding directionality to the edges, and adding normalized edge weights based on the number of interactions. For the chosen 7,385 tweets, we collect 100,302 retweets spanning them, resulting in a temporal edge list of size 99,428 where nodes are users and edges are retweets.

\begin{figure}[h]
    \includegraphics[width=\columnwidth]{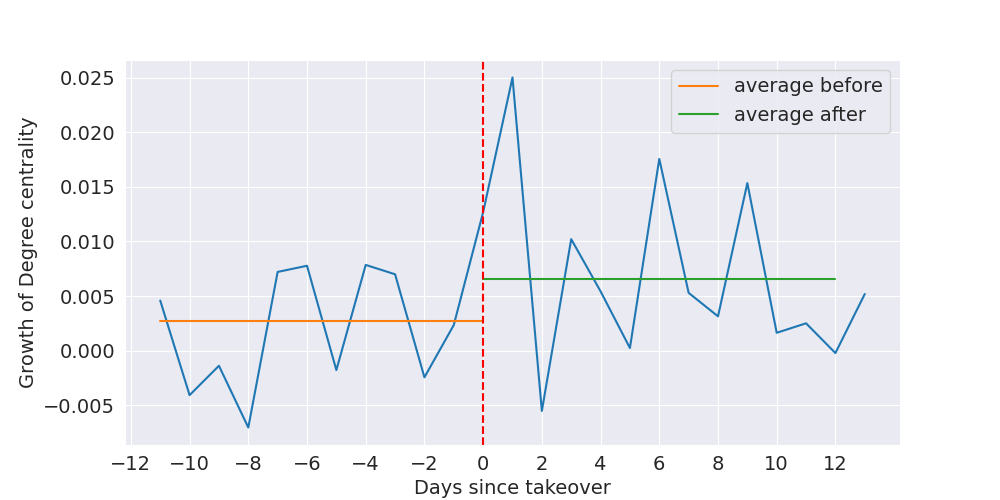}
    \caption{Rate of growth of the average degree centrality of nodes increases by 144.44\% post-takeover}
    \label{fig:degree}
\end{figure}

\begin{figure}[]
    \centering
    \includegraphics[width=\columnwidth]{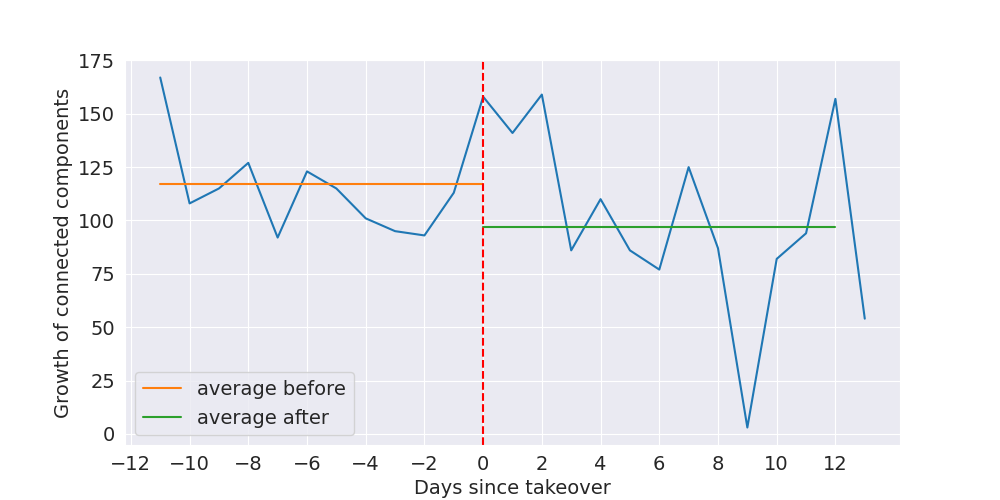}
    \caption{Rate of growth of the number of connected components decreases by 17.3\% post-takeover}
    \label{fig:ccomps}
\end{figure}

Similar to \citet{auditing}, we observe an average frequency of hateful tweets increase from 15,337 tweets per day to 16,658 after the relaxation. Examining the retweet network's temporal evolution manually, we find that the hate community's structure evolves a lot internally and also in its interaction with the rest of the network. The network expands primarily through bridge nodes while some communities grow within themselves. We observe new cliques forming as well as existing cliques merging. The initial network is visualized in Figure \ref{fig:start}, and a subgraph of the same size sampled from the final day with the same amount of nodes is visualized in Figure \ref{fig:final}, where we can notice the interactions becoming denser and communities merging.

The average edge influx per day of the network increases after the relaxation from 1,793 to 4,814 (168\% increase), suggesting a sudden rise in the activeness in the user communities. The average growth rate of the degree of nodes (note that we are talking about the derivative of increase) also increases from $2.7e-3$ to $6.6e-3$ (144\% increase) after the relaxation, as seen in Figure \ref{fig:degree}. Interestingly, despite this evident growth in network size and interactions, the average growth rate of distinct connected components decreases from 117 to 97 (17\% decrease), as shown in Figure~\ref{fig:ccomps}. This counter-intuitive trend hints at the potential merging of previously separate communities and the emergence of influential bridge users facilitating the flow of information across different segments of the network.

These findings indicate that following the relaxation, there is not only an increase in hate speech but also a rise in the engagement and propagation of such content across the platform.

\subsection{Can we (early) detect the users who drove the change in this landscape?}
\label{sec:rq3}
Identifying influencers in a time-evolving network can give insights into which communities drive the change and which users lead them. We experiment with various methods and exploit both the network information and the tweets themselves to identify the set of most influential users in the hate network.

\begin{table}[]
\centering
\caption{Representative words in the user bios of top-ranked users by MPR}
\label{tab:bio-words}
\begin{tabular}{|c|c|}
\hline
\textbf{Keyword}             & \textbf{Log-Odds Score $(1e-2)$} \\
\hline 
MAGA                     & 2.536\\
\hline 
Gaslighting                     & 2.203\\
\hline 
Self Governance             & 1.859 \\
\hline 
ACAB                      & 1.747\\
\hline 
Biden                          & 1.166\\
\hline
Prochoice                          & 0.820\\
\hline
Anti Communist                          & 0.269\\
\hline
\end{tabular}
\end{table}

\begin{figure*}[]
     \centering
     \begin{subfigure}{.47\textwidth}
         \centering
         \includegraphics[width=\linewidth]{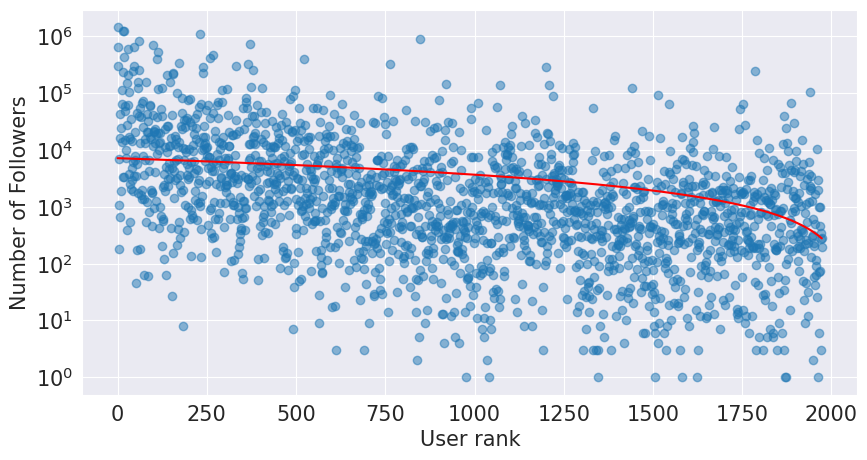}
         \caption{Number of followers $(\rho = -0.429)$}
         \label{fig:mpr1}
     \end{subfigure}
     \hspace{5mm}
     \begin{subfigure}{.47\textwidth}
         \centering
         \includegraphics[width=\linewidth]{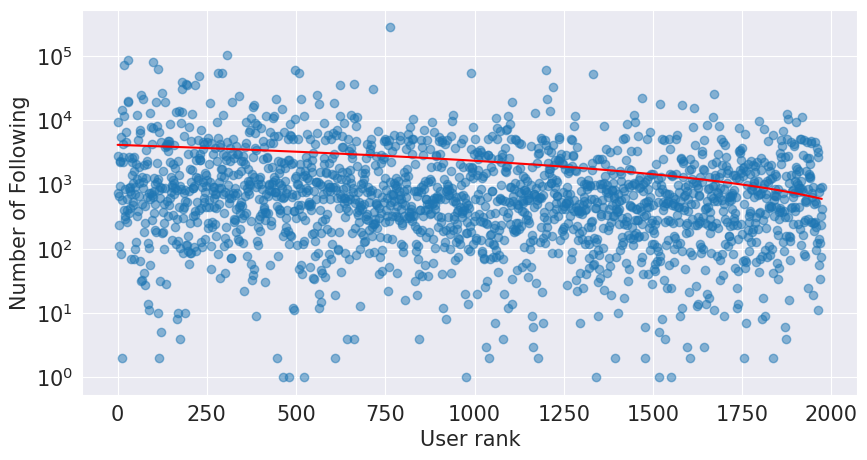}
         \caption{Number of following $(\rho = -0.138)$}
         \label{fig:mpr2}
     \end{subfigure}
    \caption{Spearman correlation $(\rho)$ between MPR rank and user profile metrics for the top 2000 users}
    \label{fig:mpr-corr}
\end{figure*}

\subsubsection{Moving PageRank (MPR)}
\label{sec:mpr}
We propose the Moving PageRank (henceforth referred to as MPR) method to identify the set of users who drive the growth of the hate interaction network. We calculate the PageRank (PR) for all the nodes at every network snapshot and then use a combination of the following three methods to find users who drive the change. In the following, $T_1$ denotes the timestep just before the takeover, $T_2$ denotes the final timestep, and $x$ denotes a user.
\begin{enumerate}[label=(\alph*)]
    \item Sum of PR change across all timestamps:
    \begin{equation*}
        f_1(x) = \enspace \sum_{t=2}^{T_2}\mid {PR}_{t}(x) - {PR}_{t-1}(x)\mid
    \end{equation*}
    \item Maximum PR change between timestamps:
    \begin{equation*}
        \hspace{-2mm} f_2(x) = \enspace {max}_{t=2}^{T_2}\mid {PR}_{t}(x) - {PR}_{t-1}(x)\mid
    \end{equation*}
    \item Maximum PR change before and after the takeover:
    \begin{equation*}
        \hspace{-6mm} f_3(x) = \mid {max}_{t=1}^{T_{1}}{PR}_{t}(x) - {max}_{t=T_{1}}^{T_{2}}{PR}_{t}(x) \mid
    \end{equation*}
\end{enumerate}
We take the intersection of sets of top 1000 users identified by $f_1, f_2$, and $f_3$ to converge on the final set,
\begin{equation}
    \mid f \mid \enspace = \enspace \mid f_1 \mid_{1000} \cap \mid f_2 \mid_{1000} \cap \mid f_3 \mid_{1000}
\end{equation}

Our method identifies 57 key nodes within the retweet network without directly attributing negative behaviors to identifiable individuals, focusing instead on these accounts' structural roles in information diffusion. We also manually verify these key users and weed out false positive accounts that crept into the set because of their popularity and the keywords used. Similar to what \citet{safak2022elon} observe, we observe a heavy right-wing presence in most of the key users detected by our methods, who vocally counter liberal culture and are often Trump allies, with a few exceptions. Moreover, the key influencers include a spectrum of political profiles, from tinfoil hat populism and sexism to aggressive MAGA rhetoric and misinformation, contrasted with pro-Biden stance and critique of right-wing hate speech. 

For the sake of user privacy, we do not perform any profile-level manual qualitative analysis. We rather analyze the bios of the top users collectively and find that most of the profiles indicate their political stances and ideologies.

As shown in Table \ref{tab:bio-words}, the presence of keywords like `MAGA', `Anti Communist', and `Self Governance' suggests a strong presence of right-wing, conservative, and potentially extremist viewpoints among these influential users. On the other hand, keywords like `Prochoice' and `Biden' indicate the existence of liberal or left-leaning voices as well, though with lower log-odds scores. The occurrence of terms like `Gaslighting' and `ACAB' (an acronym for ``All Cops Are Bastards'') points toward anti-establishment and potentially extremist ideologies. These keywords in user bios highlight the polarized political landscape and the diverse range of ideological perspectives represented among the key influencers facilitating the spread of hate speech on the platform.

\subsubsection{Early Detection of Influential Users}
MPR identifies influential users by analyzing their structural position and its evolution in the network. We investigate whether examining static user profile characteristics, such as follower/following counts and historical tweets before the takeover, could early identify key actors facilitating hate speech propagation.

\begin{table}[h]
\centering
\caption{R2 scores for Regression models trained on different feature sets for early detection}
\label{tab:regression}
\begin{tabular}{|c|c|c|c|}
\hline
\textbf{Method}              & \textbf{F1}   & \textbf{F2}   & \textbf{F1+F2} \\ \hline
Linear Regression   & 0.05 & 0.07 & 0.26  \\ \hline
AdaBoost Regression & 0.22 & 0.04 & 0.09      \\ \hline
\end{tabular}
\end{table}

We generate the first feature set (F1) containing profile metrics such as the number of followers, followings, and tweets, the age of the account, and the description length. We run the Spearman correlation $(\rho)$ \citep{spearman} test between the ranks generated by MPR and each feature and report the two highest ones. We find a correlation of -0.429 for the follower counts, while the correlation with the number of accounts a user follows is even weaker at -0.138 (Figure \ref{fig:mpr-corr}). This indicates that even the strongest correlated profile metric might not be a strong indicator.

We compile the second feature set (F2) using the mean-pooled Sentence-BERT,\footnote{\href{https://huggingface.co/sentence-transformers/all-mpnet-base-v2}{sentence-transformers/all-mpnet-base-v2}} embeddings for each user based on all their tweets, retweets, quotes, and replies before the takeover. 

To assess whether standard profile metrics and textual content alone can reliably predict MPR ranks, we train Linear and AdaBoost regression models on three combinations of these features (F1, F2, F1+F2) and report the R2 scores for each.

As shown in Table \ref{tab:regression}, even the best-performing model achieves an R2 score of only 0.26, indicating that user profile characteristics and historical tweet content alone explain just about a quarter of the variance in the MPR ranks.  
Linear Regression shows minimal improvement when switching from F1 to F2, suggesting that textual content provides slightly more predictive power than static profile metrics. 
However, combining both significantly improves performance, highlighting that user influence on hate speech diffusion is a mix of profile traits and content nature. 
Interestingly, for the AdaBoost Regression, we see contrasting results where F1 alone achieves a reasonably high R2 of 0.22, but adding F2 leads to a drastic drop in performance to 0.09. A potential explanation for this could be that AdaBoost, being an ensemble method, is able to effectively model the non-linear relationships between profile features and MPR ranks. However, when introducing high-dimensional textual embeddings, overfitting may occur, causing the model to prioritize noise over actual predictive signals from the features.

This analysis reveals that while profile metrics and historical tweets provide some signal, hate speech propagation is primarily driven by complex network effects that conventional user profile metrics and user tweets alone cannot fully capture. MPR better models these dynamics by tracking the evolving network structure and information flow over time rather than relying on static and textual data alone. For example, users with relatively few followers can still act as bridge nodes, connecting communities and facilitating hate content spread via retweets/quotes over time, gaining centrality quantified by MPR.

\section{Discussion}
Our study uncovers concerning trends following Elon Musk's Twitter takeover and subsequent relaxation of moderation standards. The findings indicate that allowing unvetted free speech facilitated an increase in hate speech targeting vulnerable communities like LGBTQ+, liberals, and ethnic minorities. Offensive terminology associated with racism, sexism, and ableism saw a sharp rise in usage across the platform (section \ref{sec:mpr}).

Our analysis (section \ref{sec:log-odds}) uncovers how the relaxation of moderation enabled a disturbing shift in the language and rhetoric used to target different communities. The increased usage of derogatory terms like `tra*ny', `schizo', and racist slurs signals a bleak regression towards more aggressive and explicit forms of hate speech. This deterioration of content points to how uncontrolled free speech can provide cover for the normalization of hate under the disguise of openness. The heightened discrimination against groups like the LGBTQ+ community and ethnic minorities through such language can incite further hostility and marginalization in the offline world and cause severe psychological impacts on people \citep{psych}. Loosening restrictions can rapidly alter linguistic norms and the boundaries of what speech gets visibility on digital landscapes. Proactive counter-speech campaigns to elevate civil, inclusive rhetoric may be necessary countermeasures.

The semantic analysis reveals how discussions around content moderation policies, free speech principles, and hate speech became increasingly intertwined post-relaxation (section \ref{sec:w2v}). Paradoxically, the push for liberal speech norms appeared to embolden voices fundamentally opposed to such freedoms. Political polarization was also catalyzed, with liberals facing intensifying targeting through far-right rhetoric and derogatory terminology.

Analysis of the hate interaction network exposed the emergence of tightly-knit communities joined by bridge users disseminating hateful content (section \ref{sec:rq2}). The surge in interactions between previously disparate groups merging into larger hateful clusters points to an escalating propagation of such toxic views enabled by the moderation changes.

Identification of influential actors driving these network dynamics (section \ref{sec:rq3}) reveals many are self-acknowledged far-right voices with records of promoting misinformation, sexism, anti-immigrant stances, and false claims of election rigging. The list also features anti-Trump voices, reflecting the nuanced landscape. We also find that only the profile metrics and the linguistic insights from user tweets are insufficient to identify users selected by MPR, hinting at the paramount importance of studying network evolution.

One practical application of our methodology could be to stagger the relaxation of content moderation policies for identified influential users. By pinpointing the few key individuals contributing disproportionately to the surge in hate speech after moderation is loosened, platforms could delay extending such policy relaxations to these actors. This measured approach could help mitigate the rapid proliferation of hate speech enabled by influential provocateurs.

Our findings echo previous research on platforms embracing unrestrictive speech policies, such as the analysis of Gab \citep{gab}, which found it quickly became an insulated ecosystem overrun by extreme right-wing ideology, hate speech, and conspiracies due to minimal moderation. We observe similar phenomena on Twitter - the merging of hateful communities facilitated by influential users upon relaxing content moderation. These findings highlight the need for balanced platform governance that preserves open discourse while countering abuse and misinformation. However, we acknowledge the complexities of balancing free speech with effective moderation. Unfettered speech freedom enables diverse viewpoints but risks enabling the unchecked spread of harmful rhetoric. We propose leveraging counter-speech measures and credible counter-narratives \citep{counter}, transparent community-driven policies and alternative moderation approaches like user-driven systems \citep{mod} like Community Notes or AI assistance with human-in-the-loop. These strategies must also account for contextual and cultural nuances in interpreting hate speech across societal norms \citep{understanding,mixedm}. By adopting nuanced, adaptive approaches, platforms can foster inclusive spaces while upholding free expression principles without providing ideological extremists freedom to proliferate harmful content.

\section{Conclusion}
We examine how the relaxation of moderation on Twitter after Elon Musk's takeover affects the platform's interaction dynamics and its users. We observe that the relaxation catalyzes the increase of hate speech against most of the commonly targeted communities and, ironically, against the promotion of free speech as well. They also set the stage for targeted political hate against their opposition. Our findings illuminate the critical need for social media platforms to balance free speech with effective moderation strategies by employing counteractive measures (like Community Notes). 
We hope that future works explore proactive measures that can be implemented to foster healthy online discourses without infringing on user freedoms. 

\noindent \textbf{Ethical statement.} In our work, we have exclusively used publicly available tweets collected via Twitter's Academic API, designed for research purposes. Despite the public nature of this data, we recognize the ethical obligation to preserve the anonymity and privacy of individuals. It is also crucial to highlight that our annotation process was designed to be user identity-agnostic, with annotators being shielded from any personal information about users to prevent potential biases. Therefore, all data has been anonymized in our analysis, with no direct quotations or identifiable information such as profile metrics being used in our analysis.

\section{Limitations}
While our study provides valuable insights into the impact of relaxed moderation on hate speech dynamics, we acknowledge potential limitations. The first is the bias that may be induced due to the keyword selection, for which we try our best to keep it balanced and best representative of a wide range of interests.

The second limitation of our study is the inability to establish a clear causal link between Elon Musk's takeover of Twitter and relaxed content moderation policies as the sole driver of increased hate speech on the platform. The sociopolitical environment surrounding the new ownership and Musk's publicly stated reasons for the takeover could have independently influenced certain user behaviors, regardless of concrete policy changes. The effects we observed could potentially correlate with, rather than directly resulting from, the new moderation approach. Moreover, it is inherently difficult to separate the relaxed moderation from confounding factors like news cycles, public discourse, and perceived changes in platform that simultaneously shifted during the transition period. Although our analysis accounts for some of these factors, completely isolating the policy impact through a hypothetical scenario is infeasible.

Categorizing users as hate perpetrators based solely on algorithmic outputs, without human validation, can raise ethical concerns about potential mischaracterization or unfair targeting. We also recognize that any form of user labeling, even if anonymized, should be undertaken with caution and transparency. Ideally, such methods should involve a human-in-the-loop process to mitigate erroneous classifications. While we can not guarantee the generalizability of our findings to other platforms, we hope that it serves as a primer for motivating necessary precautionary measures.

\bibliography{main}



\end{document}